# Geospatial Metadata Improves Discoverability by Connecting Datasets Across Scientific Disciplines

Daniel Ebanks and Devika Jain*
Institute for Quantitative Social Science, Harvard University
Center for Geographic Analysis, Harvard University

**Corresponding Author*

*Email address:* *kakkar@fas.harvard.edu* *(Devika Jain)*

## Abstract

Research data repositories are essential infrastructure for scientific inquiry and for ensuring research datasets follow FAIR (Findable, Accessible, Interoperable, and Reusable) principles. At the same time the effectiveness of repositories at promoting reuse depends on the quality and completeness of geospatial and thematic metadata, which is usually supplied voluntarily by a data depositor or researcher. Yet, due to the voluntary nature of such user-derived metadata, curation and the resource constraints facing research data repositories, it is no surprise that even on the Harvard Dataverse Repository, the world's largest general-purpose research repository, most metadata fields are missing. A missing metadata field represents a loss of information, so an incomplete metadata record is also a less interoperable record. In fact, we find that datasets with more missing metadata are associated with fewer downstream citations, and, more to the point of this paper, with fewer resolvable connections to other datasets. The implications for geospatial datasets are even more stark, as only 0.3 percent of research datasets have a bounding box, nearly all such bounding boxes are archival data points rather than full geographic shapes. We find that the value in geospatial metadata is that location helps connect datasets' concepts across disciplines. After embedding all of the datasets on the Harvard Dataverse into a metadata knowledge graph, we find that datasets are twice as likely to be connected across scientific disciplines through shared geospatial metadata than through keyword metadata, evidence that geographic metadata is the most reliable attribute we have for making datasets interoperable across the disciplinary boundaries that keyword vocabularies do not cross. We train a small language model and fine-tune it on datasets that are stored on the Harvard Dataverse repository. We find that through geospatial metadata enrichment, we are able to increase the share of datasets from different disciplines that are connected via metadata elements from 58.5 to 63.2 percent.

## 1. Introduction

Research data repositories both store and encode vital information, both in their underlying datasets and their associated metadata, about research data that makes them findable, accessible, interoperable, and reusable to researchers and scientists engaged in scientific inquiry at the frontier of research knowledge. In this paper, we look to use LLM tools to enrich metadata elements for the largest such general data repository in the world, the Harvard Dataverse Repository, and for the first time at scale. Home to over 150,000 research datasets and millions of individual data files, open research repositories rely on depositors to provide detailed metadata and descriptions about their data. In this paper, we fine-tune a small open-weight LLM to predict keyword metadata, and we evaluate it out of sample against the keywords that depositors supplied. We then apply the same treatment to geographic coverage, and we report both experiments in Section 5. Variable-level metadata is the one extension we propose without yet reporting, and Section 6 sets out that design. By enriching these metadata with validated LLM approaches, we aim to provide the most useful information available to researchers so that they can find data most suited to answer their research questions, especially where such datasets already exist but are difficult for researchers to identify.

A research data repository serves a key role as the record of account for a dataset in the research data lifecycle. Complete metadata helps repositories promote reuse and downstream scientific impact for these datasets because metadata helps potential reusers assess if a dataset is fit to help answer their scientific questions. Additionally, a repository differs in its function from a file server in that it has a fiduciary responsibility to serve as a record of account for the research dataset. Repositories have a responsibility to preserve data for future reuse and serve as a canonical record for the data. On the other hand, a file server's primary function is for users to store and retrieve files, but users have no expectation that a file server will preserve and record these files or that there is any permanent character associated with the files on the server. A research data repository also supplies persistent identifiers for datasets, as well as records dataset version histories. These repositories often steward data as well, and potentially provide permissioning controls, requiring structured descriptions of the data and its provenance, and recording  related publications that are derived from a dataset, all in the service of keeping the data interpretable findable, and reusable even long after the original researchers who created the data have stopped maintaining the dataset (King 2007; Crosas 2011; Crosas et al. 2015).

This metadata provides valuable information that helps researchers go from the facts they know to learn new scientific findings they do not. Without a detailed description and relatively complete metadata, a dataset is findable only by someone who already knows it exists, which defeats the purpose of depositing it in an open repository.

Consistent with the literature on the value of metadata quality for dataset reuse, we find strong associational evidence that increased metadata completeness is correlated with increased downstream citations. Obviously, citations and metadata quality are going to be confounded with overall research quality and researcher diligence (just two obvious confounders), but we believe that improving geospatial metadata completeness will help improve discoverability, as evidenced by the cross-disciplinary connections we find when we embed the data in a knowledge graph.

A substantial literature connects metadata to improved rates of and quality of scientific data reuse (Park 2009; Chapman et al. 2020), and our results extend on these foundations in two ways. It has been long noted in the literature that the open sharing of research artifacts is associated with an increased rate of downstream citation (Piwowar and Vision 2013; Colavizza et al. 2020), curation and documentation predict which shared datasets get reused (Hemphill et al. 2022), and further, there is qualitative work showing that data reusers need dataset context, provenance, and information about dataset fitness for purpose and quality (Koesten et al. 2020; Gregory and Koesten 2022; Jiang 2024). Wang and Strong (1996) define high-quality data as data that are fit for use. Huvila et al. (2025) extend documentation to paradata, the records of how data were created and handled, and Hansson (2025) warns that metadata requirements can become "imaginary demands", a matter of compliance as a last step for a journal to publish a paper or to satisfy the regulations of a granting agency.

In this paper, we make three key contributions towards the study of the role of geospatial metadata in promoting cross-disciplinary dataset discovery.  First, we find that more complete metadata predicts more citations for a dataset and its related publications, extending known facts about open research to the research data repository case. Despite this association, because metadata provision is costly for researchers and the benefits are not immediately obvious, we find that 59.4 percent of deposits have never been cited at all. Even so, the metadata elements depositors fail to complete are the same pieces of information that are most helpful for researchers who might want to later find and use the data (Hemphill et al. 2022). Second, we innovate on how to measure dataset discovery and fitness for use by embedding our data in a knowledge graph that includes related publications, topic, subject, and geographic extent. We can then show that data with geospatial metadata is twice as likely to connect datasets across disciplines than those with keywords alone. These results are suggestive that data with high quality geospatial metadata therefore promotes more discovery of seemingly unrelated datasets, the type of reuse that helps promote the most creative research. Finally, we show that a small fine-tuned language model can predict and enrich metadata elements, further improving our ability to uncover seemingly unrelated datasets. Despite this critical role played by geospatial metadata in helping promote discovery, we also find that geospatial metadata is the most incomplete on the Harvard Dataverse. This suggests that geospatial metadata enrichment is potentially a low cost and scalable mechanism by which to help promote dataset reuse across the many disciplines represented in the Harvard Dataverse, including astronomy, arts and humanities, social sciences, public health, economics, and the biomedical sciences.

We organize the remainder of the article as follows. Section 2 describes how we collected the dataset of metadata, the citations for datasets, and constructed the knowledge graph. Section 3 motivates the knowledge graph as a tool for measuring how datasets connect across scientific disciplines. Section 4 defines how we measure metadata missingness, summary statistics on the knowledge graph and the encoded relationships between datasets, the models we used for metadata enrichment, and the frameworks we employed for evaluation and validation. Section 5 reports the results related to citations and metadata completeness, the graph analysis, the geospatial analysis, and the enrichment experimental results. Section 6 discusses limitations, and Section 7 concludes and offers future paths of research.

## 2. Data: Citations are a proxy for downstream reuse

Our data come from the 2025 published dataset of the Harvard Dataverse Repository metadata for all deposited datasets. Datasets are collections of files that are associated with a dataset, and can include code, tabular data, image data, readme's, documents, json files, and many others. We analyze 102,650 published, non-restricted datasets and their associated metadata. Each record includes the metadata that is provided by researchers upon deposit. This data is collected in the repository software at time of submission. Elements of the metadata include title, description, subject, keywords, topic classification, related publications, geographic coverage, the spatial unit of observation, bounding box coordinates, license, grant information, and author identifiers.

In terms of geospatial coverage, a data depositor provides information such as the country, a state, a city, or an informal region name into the geospatial metadata block. We therefore treat a location as the normalized string a depositor typed and two datasets are recorded as sharing a location only when these geospatial text strings are in agreement after that normalization. We are intentionally conservative in our approach, as we undercounts shared geography wherever depositors provide different spellings for the same place-name. We target this problem for enrichment, as we describe in Section 6. We also have access to the file level data of each dataset, so we can also analyze documentation such as README files, codebooks, and replication code to further train our enrichment model. We record the completeness of thirteen metadata fields which are relevant for dataset discovery, and we report the rate of missingness in Table 1. We find that 91.1% of subjects are complete, while only 8.7% of datasets have a readme, and fewer still have a codebook or data dictionary file. We also see that 43.5% of datasets contain some geographic information.

**Table 1. Completeness by Metadata Field**

| **Field** | **Percent Complete** |
|---|---:|
| Subject | 91.1% |
| Keywords | 62.8% |
| License | 56.9% |
| Geographic location | 43.5% |
| Description of 200 characters or more | 37.2% |
| Related publication | 32.0% |
| Grant information | 30.0% |
| Variable-level metadata (UNF or DDI) | 28.0% |
| ORCID for at least one author | 21.7% |
| Replication code | 20.4% |
| Topic | 9.0% |
| README | 8.7% |
| Codebook or data dictionary file | 5.8% |

Additionally, to measure downstream research reuse, we collect citation count data from three open source APIs. First, we record direct citations to a dataset DOI from DataCite. Then, we find citations to related publications from Crossref and OpenAlex (Priem, Piwowar, and Orr 2022), but these citations are only available for datasets which either include a related publication in the related publication metadata field or which mention a related dataset in the description or title. We then record this downstream citation to a scholarly work derived from this dataset as citation for a dataset in the spirit of Altman and King (2007) and Cousijn et al. (2018). We find a total of 3,876,019 unique citations into the Harvard Dataverse, of which 15,166 (0.4 percent) cite dataset DOIs directly and 3,860,853 (99.6 percent) are to related publications. We note that this is an undercount, as depositors may not include a related publication and dataset citation practices are highly heterogeneous in the sciences. To this point, we are unable to find any citation for 59.4% of datasets.

## 3. Knowledge Graphs Quantify Cross-Disciplinary Connections between Datasets

Finally, to better analyze how datasets relate across the various disciplines in Dataverse, we follow Hogan et al. (2021) and Ehrlinger and Wöß (2016) in constructing a Knowledge Graph. This is a useful way to think about dataset discovery because reuse often involves combining primary source datasets across domains. For example, a researcher studying the long-term economic costs of exposure to air pollution in grade school in New York City would need access to geospatial and remote sensing data for air pollution, administrative records for school attendance and performance, public health data for health outcomes.

The fact these seemingly unrelated datasets are highly relevant is not ex ante obvious until we know the research question. Thus, the relational nature of a knowledge graph will then allow us to better assess the extent to which datasets are related across disciplines. Subject and topic metadata will help draw these connections. The fact that the same related paper cites four datasets is suggestive of their use. That two datasets share a geographic extent is suggestive they can be joined and related. Thus a knowledge graph representation of the data is useful because we can record many possible relationships between datasets which share metadata elements. Then we can assess which relationships (say, geospatial or topical relationships) connect datasets across disciplines.

To construct the knowledge graph, we represent the Dataverse as a labeled, directed network whose nodes represent entities such as datasets, keywords, publications, geographic locations, subjects, and academic journals. Then the edges in this graph represent typed relations between these various entities. From the dataset of Harvard Dataverse metadata, we built a graph of 215,985 nodes and 528,003 edges, of which 102,650 nodes are datasets, 56,956 are keywords, and 48,198 are publications, and an edge connects a dataset to a value for the metadata entries associated with that dataset. The node and edge types follow the Harvard Dataverse's metadata schema rather than selecting a formal ontology. Then, the Nodes that are denoted as datasets correspond to schema.org Dataset and DataCite Resource records, subject nodes use a controlled vocabulary of fifteen categories that is provided by the Harvard Dataverse at time of deposit. Finally, we map edges based on relations represented in our metadata, such as has_subject, has_keyword, has_location and related_publication. These relations map onto the DataCite subjects, geoLocations and relatedIdentifiers properties and their schema.org counterparts. The remaining node types are unstructured, e.g. keywords, locations and journal names are free string and text. We target these elements for further enrichment.

We conduct our enrichment experiments on the same dataset of Harvard Dataverse metadata. Previous attempts to enrich subject-classification metadata at scale trained on this data (Carammia, Iacus, and Porro, 2024). In that study, they trained on 76,100 records that paired each dataset's title and description with its depositor-assigned subjects, drawn from a controlled vocabulary specified by Harvard Dataverse. This approach predicts the subject of the dataset from fifteen categories, and evaluates performance on 32,619 held-out dataset records. Building on this work, we conduct a prediction experiment to classify keywords which are evaluated on 312 held-out datasets that together record 2,895 depositor keywords. For the geospatial analysis we extract every geographic bounding box in the dataset of Harvard Dataverse metadata's geospatial metadata block. Across 28,409 datasets that have at least one bounding box, 28,384 have valid latitudinal and longitudinal coordinates.

## 4. Methods

### *4.1 Computing Missingness and Citation Correlations*

We summarize each dataset's metadata completeness by counting the number of core metadata fields that are missing from the entry.

Then, *m* measures missingness for each dataset i. Concretely, m for dataset i is the count of the thirteen fields listed in Table 1 that depositors i did not complete upon deposit of the dataset. empty. To calculate a missingness index, we just take the simple sum over this count, where we weight Every metadata element equally. So, the missingness measure ranges from zero for a fully complete metadata record to thirteen for a dataset with no recorded metadata information. We then measure the correlation between metadata missingness and downstream citations. We report a Spearman rank correlation between *m* and the citation count *c*, because the rank correlation helps account for the heavily skewed distribution of citations (which have a long right tail and a lot of mass near zero). We note that 99.6 percent of the citations we observe are from a dataset's related publications. So some of the negative correlation is inherently mechanical, since a dataset that does not have an associated publication will not necessarily have any citation (although we also attempted to mitigate this problem by using an agent to go and find any missing related publications based on an exhaustive API search of OpenAlex and Datacite). We therefore report three further approaches to analyzing the association between citations and downstream use. Dropping the related publication field from m leaves a Spearman correlation of negative 0.32. Even when restricting the sample to the 32,802 datasets that point to at least one related publication, we find a negative correlation of 0.16. Finally, looking at the smaller sample of citations that refer directly to a dataset on DataCite citations to the dataset DOI, which does not rely on a related publication at all, still exhibits a negative correlation of 0.26. The correlation is robust to each of these checks, although it is weaker when we exclude more data. That said, if the underlying cause of the metadata missingness is researcher effort, they might have also put lower effort into the production of their data and downstream research products.

### *4.2 Graph Statistics*

We represent the datasets on the Harvard Dataverse as a knowledge graph in an undirected labeled network. In the results sections, we report the component structure implied by the metadata relation, the network density, and the PageRank centrality of the geographic and non-geographic nodes in the graph to gain a sense of whether geographic datasets are concentrated in any one discipline. In particular PageRank centrality (Page et al. 1999) is defined as

$$PR(v) = \frac{1-\alpha}{|V|} + \alpha \sum_{u \in N(v)} \frac{PR(u)}{deg(u)}, \quad \alpha = 0.85$$

where N(v) is the set of nodes that are directly connected to node v, and deg(u) counts the total number of edges of node u. To give an intuitive sense of what PageRank centrality means on a knowledge graph, imagine a researcher randomly perusing the Dataverse. If they start at a Dataset about electoral returns, and keep clicking around at random, PageRank tells you how often they will find themselves clicking back on that original dataset. If geographic datasets are strewn about different disciplines, geographic and non-geographic datasets should exhibit similar PageRank similarities.

We take a different approach to compare the value of geographic and non-geographic metadata types towards connecting two datasets across two or more disciplines. To measure cross-disciplinarity, we take every dataset connected by shared geographic location, every dataset connected by a keyword, and every related publication that points to more than two datasets, and count the distinct number of subjects (a proxy for disciplines) for the datasets sharing that metadata element. Intuitively, this means that when two datasets are connecting on the same place, say New York, or share the same keyword, say returns, and they come from two separate subjects, say astronomy or political science, we say that particular element connected datasets across disciplines

### *4.3 Enrichment Models and Evaluation Metrics*

All three enrichment experiments follow the same evaluative framework; we fine-tune a small open-weight model on the Harvard Dataverse repository's metadata entries for datasets, then evaluate the predictive accuracy on held-out metadata entries. The subject classifier of Carammia, Iacus, and Porro (2024) fine-tunes Llama-2 7B on the training set described in Section 2, and its headline metrics are the match rate (at least one predicted subject appears among the depositor's subjects) and the rate of finding an exact match to between our prediction and a depositor-provided label. Our keyword experiment fine-tunes Llama-3.2 3B Instruct to generate a set of potential keywords from dataset descriptions.

We train a Low-Rank Adaptation adapter (Hu et al. 2022) of rank 64, with a batch size of four and a maximum sequence length of 1,024 tokens, on one H200 GPU node[1]. We prompt the model with a fixed system message that asks the model to act as a research archivist expert in dataset metadata and asks for the answer as a bracketed list. The prompts and the training and inference scripts are publicly available (IQSS 2026; Dataverse Project 2026).

[1] With alpha= 32 and dropout = 0.1, over a 4-bit NF4 quantization of the base weights (Dettmers et al. 2023), for ten epochs at a learning rate of 0.0002 with the paged AdamW optimizer.At inference we give the merged model

We know that since keywords are free-form text, they are unlikely to match generated strings verbatim. Thus, our evaluation scores our predictions on semantic relatedness instead. We employ an LLM judge (Zheng et al. 2023) to compare each pair of one depositor-provided keyword against one generated keyword and return a binary relatedness score, J, which is one if the judge thinks the words are related or zero otherwise. The judge is the same fine-tuned Llama-3.2 3B[2] model with a limit of eleven generated tokens, and instructed to answer only with a single Yes or No for if the words are semantically similar. We gave the LLM a worked example and asked whether bread is related to butter. We have posted the prompts for the semantic similarity and the 44,001 per-pair verdicts it produced are in the enrichment codebase (IQSS 2026). We compute the semantic coverage of the depositor-labelled keyword vocabulary K by the predicted set G is

$$Coverage = \frac{1}{|K|} \sum_{k \in K} 1[\exists\, g \in G : J = 1]$$

and a precision-type analog that reverses the position of the two sets in the equation.

## 5. Results

### *5.1 Missing Metadata is Associated with Fewer Downstream Citations*

Metadata completeness is in practice a proxy measure of depositor effort and researcher fastidiousness rather than the scientific value of the scholarly artifact, but the first-order correlation serves as a useful check that our data are following known patterns of reuse behavior for other types of scholarly objects. Figure 1 plots citations against the count of missing fields across the full collection of datasets on the Harvard Dataverse repository. We show that mean citations fall by roughly two orders of magnitude as the number of missing metadata elements increases from one to thirteen. This is preliminary confirmatory evidence that metadata richness is associated with scientific reuse for research data. The field this paper is about sits near the bottom of that ledger. Geographic location is present on fewer than half of all deposits, below quantized to Q4 K M on llama.cpp with a 4,096-token context, a temperature of 0.01 and a limit of responses to 128 generated tokens

[2] We employ Q4 K M quantization, operating on llama.cpp with a temperature of 0.01

subject, keywords and license in Table 1, so the metadata element that does the most to connect datasets across disciplines is also one of the least often supplied fields in the repository. Of course, discovery is best facilitated when a researcher can connect seemingly unrelated datasets. We attempt to measure such relationships in the next section.

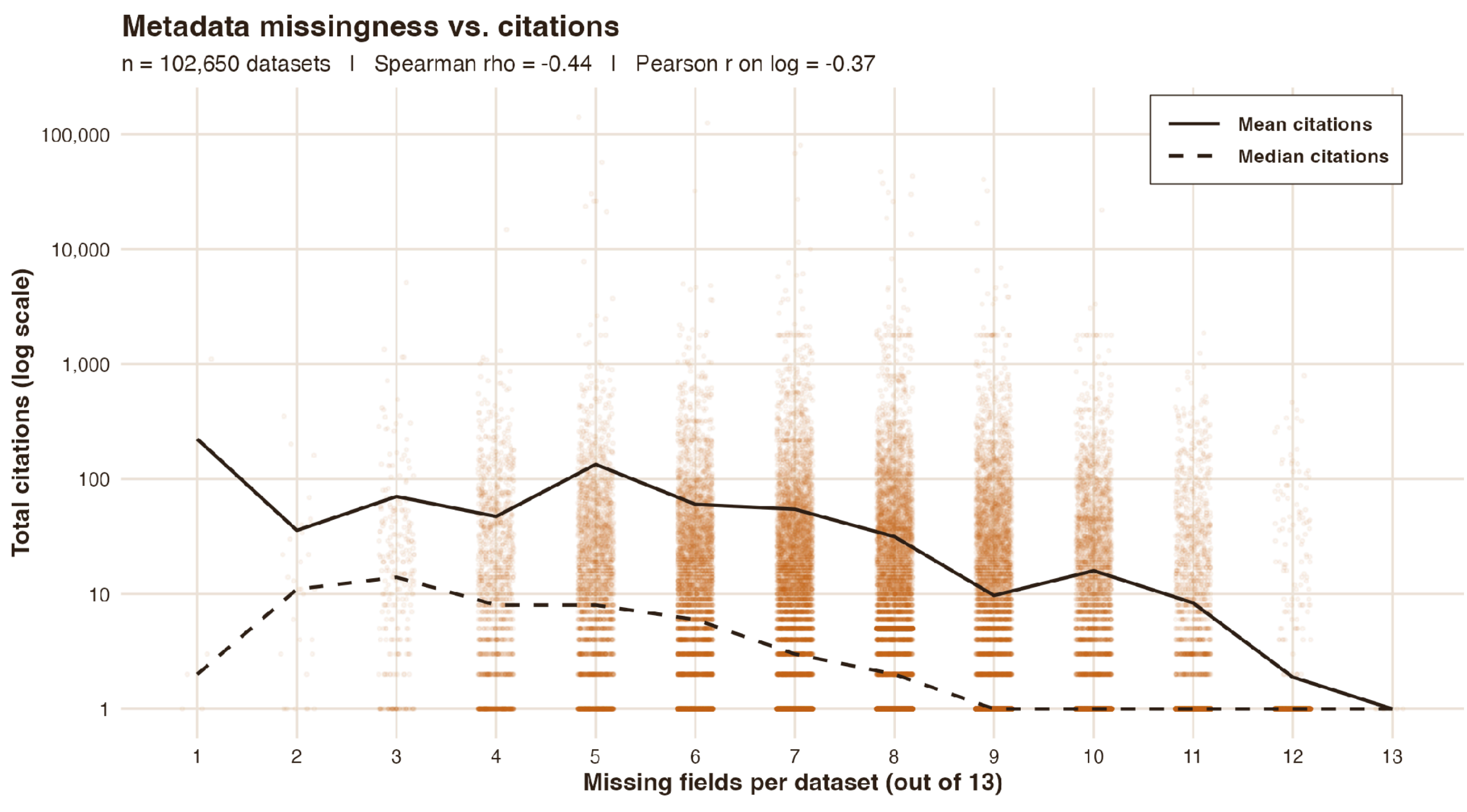

**Figure 1.** Metadata missingness and citations across 102,650 Harvard Dataverse datasets. Each point is one dataset; lines give mean (solid) and median (dashed) citations per missingness level; the vertical axis uses a log scale. Spearman rank correlation is negative 0.44.

### *5.2 Graph Structure Identifies High-Value Metadata Elements to Automatically Enrich*

We embed the Dataverse into one large relational knowledge graph. The graph topology shows a highly connected structure, with nearly all of the datasets and associated metadata elements related in one giant connected component. Recall that nodes are datasets, metadata elements (like specific subjects, topics, related papers, or geospatial values), and edges are relationships determined by a metadata schema between these various elements. So, for example, two datasets might share that they have a geographic location. The tight connections among Dataverse datasets through shared metadata elements support the goal of promoting and measuring reuse—it means our research data have many shared relationships we can exploit to better understand which datasets may seem unrelated. Overall, the knowledge graph contains 5,061 components, the largest of which is 209,315 nodes, or 96.9 percent of the total number of nodes in the graph. Any two datasets inside the giant component connect through some path of shared keywords, subjects, or publications, so given that so many paths exist, we measure dataset similarity through path length. Density is low, at 2.26 x 10^-5, the expected order for a network in which each dataset connects to a handful of attribute nodes rather than to other datasets directly. Geospatially tagged datasets average a PageRank of 4.78 x 10^-6 against 4.85 x 10^-6 for everything else, a ratio of 0.99, so geospatial metadata are equally likely to be connected to datasets across the Dataverse, and are not concentrated in any one part of the repository.

Figure 2 shows a subgraph based on the repository's most-cited datasets, and the hub-and-spoke structure that these statistics describe is visible upon inspection.

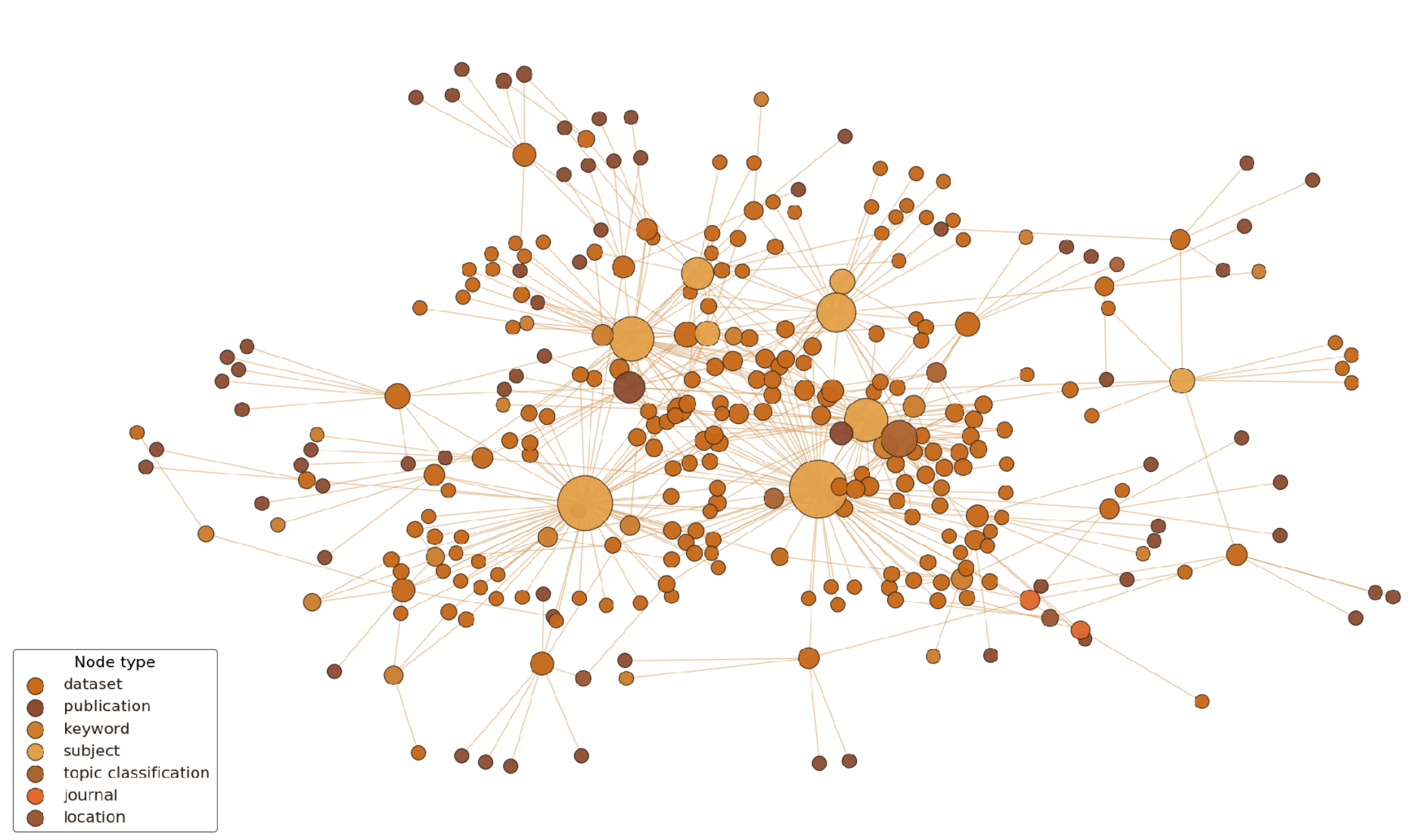

**Figure 2.** A subgraph of the knowledge graph anchored on the most-cited datasets (309 nodes, 490 edges). Node color marks entity type; node size marks degree. Datasets connect through shared subjects, keywords, publications, journals, and locations.

Shared keywords account for 207,076 of the 528,003 edges, or 39.2 percent of the total number of edges on the graph. Yet 81.8 percent of the keywords provided by data depositors appear in only one dataset, and thus are not serving the purpose of connecting datasets which may be seemingly unrelated. One potential area of metadata enrichment would be to find semantically similar keywords and collapse them into one entry. Since keywords are free text boxes, data depositors often use different words to represent the same keyword concept (e.g., "election returns," "electoral returns," "election data") that the graph treats as distinct nodes. Keyword generation from a consistent learned vocabulary would consolidate the various ways researchers phrase the same keyword concept so they can feasibly be used to connect datasets. The same fragmentation afflicts place names, and for the same reason. USA, United States, U.S. and United States of America are four nodes in our graph and one country on a map, so the cross-disciplinary reach we measure for geography in Section 5.3 is a lower bound on the true reach of geographic metadata.

Emblematic of the long-tailed nature of dataset curation effort, we report in Figure 3 the distribution of edges per dataset, with the median node having five edges and a maximum of 190. We note that datasets with the most metadata entries are a direct proxy for depositor and repository curator effort to ensure high quality data deposits. Given the current manual nature of this work, it is no surprise that only a select few datasets are given enough attention for a high quality metadata entry.

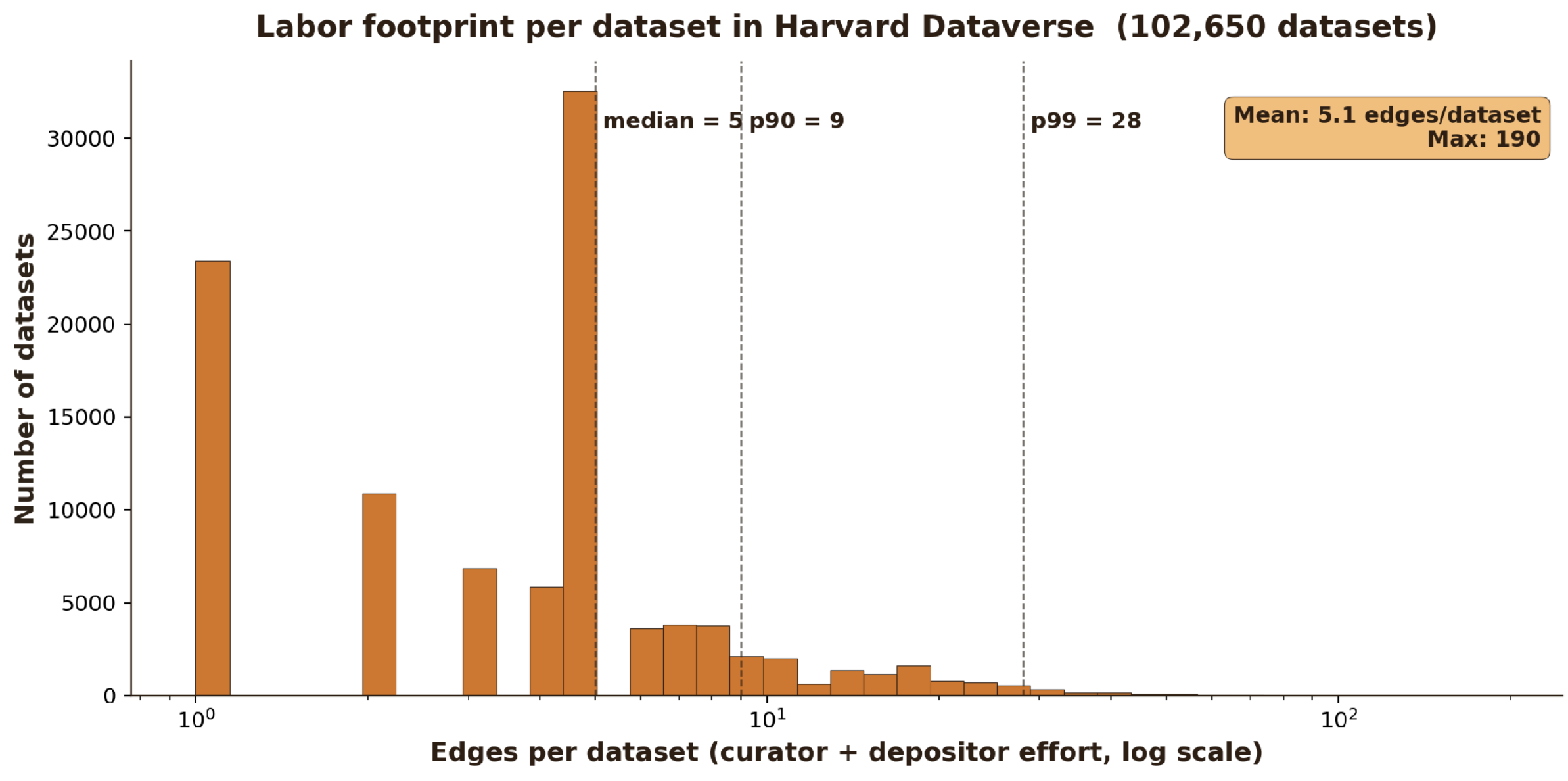

**Figure 3.** Edges per dataset in the knowledge graph, on a log scale (102,650 datasets). The median dataset holds five edges; the 90th percentile holds nine, the 99th holds 28, and the maximum holds 190.

Of the 102,650 datasets, 44,607 (43.5 percent) include at least one geospatial metadata field, leaving 56.5 percent of the datasets with no spatial information, and 72.3 percent missing bounding box coordinates.

### *5.3 Geospatial Data Connects Datasets Across Disciplines on the Harvard Dataverse*

Geospatial metadata is actually the strongest link between seemingly unrelated datasets on the Harvard Dataverse repository. Geographic features can serve as a link across disparate datasets from an array of academic disciplines, including public health, sociology, political science, economics, geology, climate, remote sensing, and others (Janowicz et al. 2020; Mai et al. 2024). At the same time, geospatial metadata on the Dataverse repository is currently sparse. Despite this sparsity, geospatial metadata, when they are available, are still reliably connected datasets across the disciplines that deposit their data in the repository. Datasets that include bounding boxes in their metadata are emblematic of the geospatial issues in Harvard Dataverse. The bounding boxes that exist for datasets on the Harvard Dataverse are usually not boxes at all, but rather points. Of the 28,384 datasets containing at least one valid bounding box, 99.4 percent enclose an area smaller than one square degree on a map. As it currently exists, the bounding box information on the Harvard Dataverse functions more as a catalog of points rather than as a description of rich geospatial information for any particular dataset[3].

Outside Arts and Humanities, 221 dataset deposits in the Harvard Dataverse include a bounding box. Figure 4 maps every geographic centroid and shows that the Ashkelon deposit predominates over the other data points.

---

[3] The boxes also belong overwhelmingly to one collection in the repository, a bulk deposit from the Leon Levy Expedition to Ashkelon. This collection of datasets has its own standardized metadata requirements and curation quality controls, closer in spirit to ISO 19115 practice (ISO 2014) than to anything else in the repository. These requirements happen to include latitude-longitude coordinates for any data artifact deposited as part of this collection. Most of these datasets' coordinates lie in the eastern Mediterranean (31.7 degrees east, 34.5 degrees north).

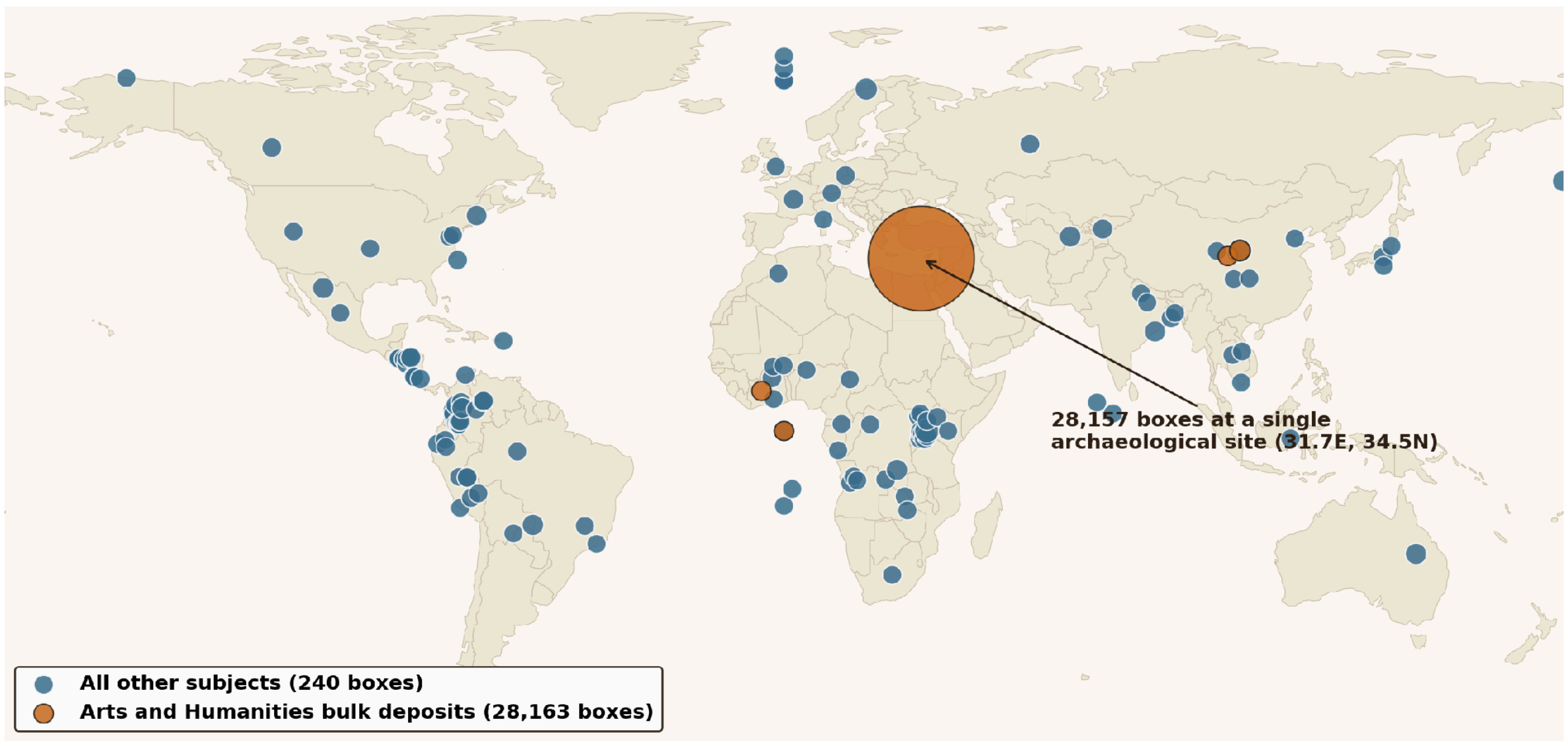

**Figure 4.** Bounding-box locations in Harvard Dataverse, with circle area scaled to the number of boxes at each location. Orange circles are Arts and Humanities bulk deposits (28,163 datasets, 28,157 of them at a single archaeological site); blue circles are all other subjects (221 datasets).

The inclusion of geographic information with a dataset deposit is correlated with some additional increase in predicted citations for some disciplines but not others. Table 2 compares by field for the datasets whose deposits include bounding boxes that encompass a geometry with positive area, with two outliers removed. We find that Earth and Environmental Science datasets with bounding boxes average 37.1 citations against 46.0 for other datasets in that field. Then, Agricultural Sciences datasets with bounding boxes average 59.2 citations, compared with 49.7 for those without. Finally, in Social Sciences, datasets with bounding boxes have 56.7 citations, compared with 40.4 citations for datasets lacking a bounding box.

**Table 2. Mean citations by field for datasets with and without a positive-area bounding box (two outlier climate surfaces excluded).**

| Field | n, with bounding box | Mean citations | n, without | Mean citations |
|---|---:|---:|---:|---:|
| Earth and Environmental Sciences | 154 | 37.1 | 4,728 | 46.0 |
| Agricultural Sciences | 82 | 59.2 | 2,908 | 49.7 |
| Social Sciences | 42 | 56.7 | 41,783 | 40.4 |

Despite mixed evidence for correlations with downstream citations, geographic metadata elements are nonetheless the most likely to connect datasets across disciplines in the Dataverse. Among attribute nodes with degree of at least two (meaning they connect to two or more datasets), 58.5 percent of nodes containing a geographic location relate datasets from two or more distinct subjects (a mean of 2.21 subjects per node), while 32.2 percent of keyword nodes do the same (a mean of 1.53). Related-publication nodes fall between the two; a publication shared by at least two datasets bridges two or more subjects 42.6 percent of the time, but only 6.4 percent of related publications connect to more than one dataset in any case, so accounting for base rates, the likelihood a paper connects two papers at all is rare. Table 3 reports the comparison of these metadata elements in full. We find that related publications and keywords have the same mean of 1.53 subjects, even though their cross-disciplinary rates of bridging differ by ten percentage points. We note that the fact the means are the same is an artifact of the data, as a publication might cross about 2.2 subjects on the occasions when it has an interdisciplinary link, but this is a rare occurrence. At the same time a keyword bridges two subjects less frequently, and reaches about 2.6 subjects in those rare cases. Averaging over the many non-bridging nodes of each type returns both types to 1.53. Location nodes bridge both more often and across a wider array of datasets, at about 3.1 subjects per bridging node.

Geographic location is a common means to connect concepts across scientific disciplines in a way that a technical vocabulary is not so easily shared. This makes substantive sense: epidemiologists, economists, and archaeologists deposit data about the same country while sharing almost no vocabulary for how they describe or communicate about a dataset. In fact, they might have many of the same uses, but because of interdisciplinary barriers in language and technical jargon, even the best language models might struggle to make a match based on semantic similarity. Geospatial data is a convenient way to overcome this obstacle, as space and time are commonly agreed upon concepts in applied research. Enriching and predicting geographic metadata elements, therefore, is likely to improve our ability to connect datasets, which is the structural basis for the fourth step of our analysis.

**Table 3. Cross-disciplinary bridging by attribute node type (nodes shared by at least two datasets). The enriched location row adds grounded model-predicted places for the datasets that had none (Section 5.6).**

| Metadata Element | n (degree ≥ 2) | Percent connecting ≥ 2 subjects | Mean # subjects |
|---|---:|---:|---:|
| Location | 1,552 | 58.5% | 2.21 |
| Location, enriched (grounded) | 2,291 | 63.2% | 2.56 |
| Related publication | 3,087 | 42.6% | 1.53 |

| Keyword | 14,538 | 32.2% | 1.53 |
|---|---|---|---|

These results suggest that geospatial metadata is valuable information for research data infrastructures, as spatial queries (e.g., a bounding box) are answered, and cross-disciplinary connections are most easily made among these semantically similar and easier-to-resolve metadata elements. While downstream citation is of course of pressing interest to repositories, the researchers who deposit there, and the funding agencies for the research, improving geospatial metadata might help improve dataset discovery even if the benefits of that metadata are not immediately obvious in the citation metrics.

### *5.4 Feasibility of Automated Enrichment*

We conducted a small proof-of-concept study to show that we can successfully enrich keyword metadata for datasets. We used a compact fine-tuned model to generate keywords that match most of the labels that depositors provided. Additionally, this approach also proposed usable keywords beyond these initial user-provided terms. For the 312 held-out datasets, the predicted set of keywords contains 4,125 unique terms (13.2 per dataset), compared with the 9.3-keyword-per-dataset average generated by a human data depositor. This result suggests that the LLM model recovers the universe of keywords generated by researchers who deposit datasets at the Harvard Dataverse, while also potentially extending and improving on this set of keywords, at only a moderate cost in precision. Of the 2,895 depositor-generated keywords, 85.0 percent have at least one semantically related keyword in the LLM-generated set. Figure 5 shows that, of the 4,125 generated keywords from our enrichment methods, 78.2 percent relate to at least one depositor-derived keyword, and for the median dataset the coverage share reaches 92.9 percent. For 96.8 percent of test datasets, the generated set contains at least one related keyword. The 21.8 percent of generated keywords with no depositor counterpart range from, upon inspection, some genuinely novel but accurate adjectives for these datasets to, of course, some generic terms that provide little marginal value for improving dataset discovery and reuse. In a future study, we would conduct expert review and labeling, which we hope will further improve keyword generation and standardization. Eventually, we will deploy these enrichments at scale for the entire Dataverse repository, with significant potential to improve downstream discovery and search since keywords drawn from one consistent vocabulary could transform otherwise isolated strings into links that connect datasets across variegated disciplines.

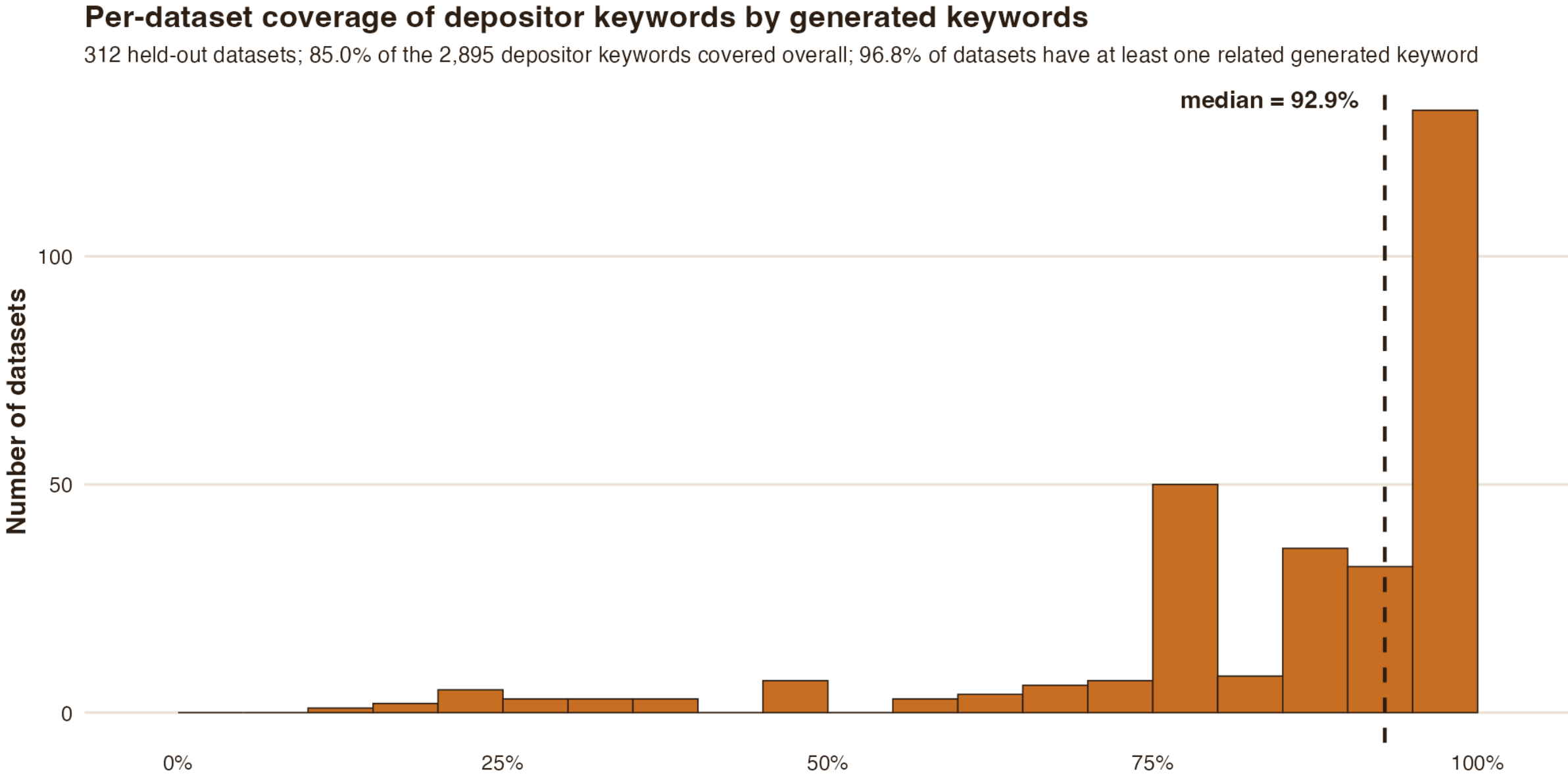

**Figure 5.** Per-dataset coverage of depositor keywords by generated keywords on the 312 held-out datasets. Each dataset contributes the share of its depositor keywords judged semantically related to at least one generated keyword; the dashed line marks the median of 92.9 percent.

### *5.5 Feasibility of Automated Subject Classification*

Notably, classification accuracy is not sensitive to the length of description, which is comforting as writing high-quality descriptions of datasets is laborious work for data depositors. With enough additional metadata and file-level data, we achieve favorable accuracy rates in matching semantic similarity for keywords. We are able to find keyword matches between our prediction and depositor-supplied keywords at a rate of 89.4 percent for descriptions under 50 characters, then 84.7 percent for descriptions that have a length of 50 to 99 characters. For descriptions that exceed 100 characters, we find that the semantic match rate between predicted keywords and LLM-generated keywords ranges from 91.2 to 97.7 percent. Notably, description length is more correlated with an increasing likelihood of finding a semantic match in this experiment. These results imply that a benefit of our approach is that it does not require detailed descriptions that may or may not be available *ex post* after data is deposited in a repository.

Most of the remaining predictive error is that our model tends to be more fastidious in labelling datasets than many of these datasets' human authors. Of the 223 datasets in our test set whose data depositors only provided the keyword "Other", our model tends to provide more information than was supplied by the author of the dataset. Our approach tends to result in a substantively meaningful and a keyword that accurately describes the dataset. Comparing across disciplines, our model correctly classifies 95 of datasets as belonging to Medicine, Health and Life Sciences, and identifies 68 datasets as Social Sciences, while 30 datasets are classified as Arts and Humanities, and 26 datasets are labeled as Computer and Information Science. Thus, an open-source language model further facilitates metadata enrichment by proposing substantive keyword classifications even in cases when the depositor failed to do so.

### *5.6 Geospatial Enrichment*

We extend the same enrichment framework for keywords to geographic metadata enrichment. We fine-tune a Llama-3.2 3 billion parameter model using metadata elements that are already available in the Harvard Dataverse repository. These include the dataset's title, description and subject. Our initial analysis is focused on the 14,838 datasets whose metadata records include at least one real geographic place. Of 1,500 held-out test datasets, the fine-tuned model finds at least one of the named geographic locations (as identified by a depositor in the dataset metadata) for 89.6 percent of datasets as opposed to only 76.8 percent for the same model before fine-tuning.

The fine-tuned model correctly predicts 66.6 percent of all depositor-provided geographic locations and has a precision of 75.2 percent. We note the largest gains from enrichment on datasets whose dataset authors supplied no subject for their dataset at time of deposit. Thus the two-step enrichment process proposed here is successfully jointly predicting missing metadata elements across multiple metadata fields at once. These are precisely the datasets where such enrichment is most helpful, as the share of datasets with geospatial resolution increases from 31.1 to 83.9 percent. Our predictions have some uncertainty, as a predicted location might vary by degree (NYC could mean its precise municipal boundaries, but if we include only Manhattan, most people understand this geographic extent to equate to New York City), we also geocode each prediction and measure its distance to the centroid of the depositor's bounding box where one is provided (Gritta et al. 2018). The median error falls from 503 to 405 kilometers, and the share within 200 kilometers rises from 26.9 to 34.6 percent. We caution there are two key limitations. First, depositor provided information is noisy ground truth, with unit strings and vague regions typed into the field, and on the boxed subset the fine-tuned model prefers country names to the watershed and site names that Earth and Environmental depositors use, so it scores lower on string match there while landing closer on the map.

We now show that geographic enrichment improves cross-disciplinary dataset discovery. We apply the fine-tuned model to the 58,154 datasets whose depositors provided no place names and report a geospatial prediction only if a place name or an analogous alias appears in the dataset's title or description. Under this convention, we conduct geospatial enrichment on 18,311 datasets. Our enrichment approach increases the share of datasets with a geographic location from 43.5 to 61.2 percent. When we rebuild the Knowledge Graph using this enriched data,, the share of location nodes that connect two or more subjects from 58.5 to 63.2 percent, and the mean number of subjects per node from 2.21 to 2.56, so enriched geography connects more datasets across more disciplines than the depositor-supplied metadata does on its own.

## 6. Discussion and proposed variable-level metadata enrichment

Metadata enrichment improves the ability to represent relations between datasets on the Harvard Dataverse in a knowledge graph. If our modeling approach can successfully predict a dataset's subject and resolve a bounding box for its geographic extent, we can represent this relational information as an edge in our knowledge graph. With enough time, we could then conduct a long-term follow-up study that isolates the true downstream effect of improved geospatial metadata quality vis-à-vis other potential metadata enrichments. For example, the share of keywords with only one dataset using that keyword is a key area that could be improved; the 43.5 percent geospatial coverage rate and the component structure are all quantities that enrichment could potentially improve. By enriching this data, we can run a simple before-and-after test of the enrichment's usefulness. As we embedded the knowledge graph and measured citations for related publications before fully enriching the missing metadata, we can measure citations and report graph elements again in a follow-up study.

We expect that in such a long-term study of data reuse on the Harvard Dataverse, the share of keywords that appear on only one dataset should fall, as the share of datasets with geographic metadata should increase from its current 43.5 percent, we should also see an increased probability of two seemingly unrelated datasets being connected in our knowledge graph representation of the Harvard Dataverse. We can then report these potentially related datasets via search or a refined "Data For You" algorithm that is tailored to the specific scientific questions of the researcher searching for secondary source data. The citation analysis from Section 5.1 is suggestive of one further statistical test of reuse. We know that datasets with more complete metadata are associated with more citations. However, we can conduct a direct quasi-causal test: if completing the metadata genuinely helps researchers find data, then the datasets we enrich should begin to catch up in citations despite the fact that the depositors were less conscientious in their deposit. This would be a test of the direct effect of metadata quality that completely abstracts away from depositor effort.

Another key observation is that geographic metadata fields are actually most likely to be missing in the disciplines whose research designs are most spatial, a pattern we did not expect before conducting this study. Medicine, Health and Life Sciences datasets only have a geographic location for only 4.4 percent of 10,093 datasets in that domain. Earth and Environmental Sciences only have geography for 14.4 percent of the 4,886 in that discipline. Business and Management have geospatial metadata for only 1.4 percent of the 2,579 in that field. Social Sciences have more comprehensive geographic coverage, with 29.4 percent of its 41,827 datasets possessing geographic metadata information. Arts and Humanities performs best, with 93.3 percent of 31,590, but this is a result of the highly curated data that is deposited by the Leon Levy Expedition to Ashkelon.[4]

In terms of future research, we aim to resolve variable-level metadata at scale for the dataset deposits in Harvard Dataverse, particularly tabular data. To the best of our knowledge, there have been few attempts by research data repositories to extract variable-level metadata at scale. Thanks to increasingly large LLM context windows and agentic approaches, we propose to load and study every variable from tabular data in our dataset to provide a fine-grained enrichment at a previously infeasible scale. As LLMs can respond and flag idiosyncrasies in real time, this enables a human researcher to make quick progress in annotating data at scale without having to review millions of files manually. The researcher will only be flagged for cases that necessitate intervention, drastically reducing the amount of labor needed to systematically extract this information in an accurate way.

For data infrastructures beyond Harvard Dataverse with geospatial data, the key practical result from this work is that semantic annotation at scale requires far less labor-intensive manual curation for a research data repository, with potentially high upside for facilitating downstream data reuse, especially if enriching geospatial metadata elements. That is, the approaches are feasible at scale even for repositories that do not have the same resources as Harvard's Dataverse. The fine-tuning frameworks use information that is already owned and maintained by the repository, so they can use supervised learning approaches for enrichments. These approaches make it plausible to validate the results and ensure they are providing accurate information. Moreover, the models we employ are open-weight and small enough to train on non-commercial-grade computational hardware. Geospatial data infrastructures could also benefit from future enrichments we could offer to a repository to convert free-form geographic text into formal bounding boxes at scale (Wilkinson et al. 2016). The same enrichment framework is transferable to any data repository and any metadata element with enough complete additional metadata for fine-tuning and validation; implementing such an enrichment is generalizable to other research data repositories with even minimal access to computing resources .

[4] The Leon Levy Expedition to Ashkelon was a project of the Harvard Museum of the Ancient Near East, directed by Lawrence E. Stager and later co-directed by Daniel M. Master, which excavated at Ashkelon from 1985 to 2016. Its Harvard Dataverse collection deposits each excavation photograph and drawing as a separate dataset under a standardized cataloguing scheme that records a latitude-longitude pair for the site, which is what produces this outlier (Leon Levy Expedition to Ashkelon 2026).

## 7. Conclusions

In this paper, we have found that datasets on the Harvard Dataverse are most likely to be connected across disciplines when they contain geospatial data. This has implications for dataset reuse and discovery, as researchers are ever more likely to engage in interdisciplinary research in the age of A.I. agents. By enriching this data at scale, we can help human researchers and their agents better discover the datasets that are fit for the purposes of answering their specific scientific questions.

One potential area of future metadata enrichment research would be to find the predictive accuracy of modeling semantically similar keywords and collapsing them into one entry to improve how well we discover which datasets are connected to the repository.

Additional future work should look to resolve geospatial records at the variable level and push georeferencing to as granular a level as possible. By making use of the variable-level data in the deposited datasets on the Dataverse, we can use the information encoded in the uploaded tabular files in a dataset to record our variable-level data, so we can extract geographies and other geospatial information that depositors directly supplied at the time of deposit.  Furthermore, the fact that missing metadata is negatively correlated with citations is suggestive of a future experimental design. If completed metadata raises discovery, the difference in citations gradient across metadata completeness should flatten for enriched datasets, a testable implication we intend to pursue in future work.

## Acknowledgments

This work was funded by the Harvard Data Science Initiative (HDSI) through HDSI Special Projects Fund 2025. We would like to thank the IQSS Dataverse curation and engineering teams for infrastructure support.

## Authors Contributions

DCE - Ideation, writing, editing, model development

DJ - Model engineering, writing, editing, and model development, and geospatial analysis

## Ethics Statement

The authors declare no competing interests.